\documentclass[preprint2]{aastex}

\newcommand{\Msun}{\mbox{$M_{\odot}$}}
\newcommand{\Rsun}{\mbox{$R_{\odot}$}}

\newcommand{\kms}{\mbox{km s$^{-1}$}}

\shorttitle{The Double K Dwarf System: 2MASS J05162281+2607387}
\shortauthors{Bayless A. J. and Orosz J. A.}

\begin{document}

\title{2MASS J05162881+2607387: A New Low-Mass Double-Lined 
Eclipsing Binary\protect\footnote{Based on 
observations obtained with the Hobby-Eberly Telescope, 
which is a joint project of the University of 
Texas at Austin, the Pennsylvania State University, 
Stanford University, 
Ludwig-Maximilians-Universit\"at M\"unchen, and 
Georg-August-Universit\"at G\"ttingen.}}

\author{Amanda J. Bayless}

\and 

\author{Jerome A. Orosz}

\affil{Department of Astronomy, San Diego State University,
5500 Campanile Drive, San Diego, CA 82182-1221}
\email{baylessa@sciences.sdsu.edu,orosz@sciences.sdsu.edu}

\begin{abstract}

We show that the star known as 2MASS J05162881+2607387 (hereafter
J0516) is a double-lined eclipsing binary with nearly identical
low-mass components.  The spectroscopic elements derived from 18
spectra obtained with the High Resolution Spectrograph on the
Hobby-Eberly Telescope during the Fall of 2005 are $K_1=88.45\pm 0.48$
km s$^{-1}$ and $K_2=90.43\pm 0.60$ km s$^{-1}$, resulting in a mass
ratio of $q=K_1/K_2= 0.978 \pm 0.018$ and minimum masses of
$M_1\sin^{3}i=0.775\pm 0.016\,M_{\odot}$ and $M_2\sin^{3}i=0.759\pm
0.012\,M_{\odot}$, respectively.  We have extensive differential
photometry of J0516 obtained over several nights between 2004
January-March (epoch 1) and 2004 October-2005 January plus 2006
January (epoch 2) using the 1m telescope at the Mount Laguna
Observatory.  The source was roughly 0.1 mag brighter in all
three bandpasses during epoch 1 when compared to epoch 2.  Also,
phased light curves from epoch 1 show considerable out-of-eclipse
variability, presumably due to bright spots on one or both stars.  In
contrast, the phased light curves from epoch 2 show little
out-of-eclipse variability.  The light curves from epoch 2 and the
radial velocity curves were analyzed using our ELC code with updated
model atmospheres for low-mass stars.  We find the following:
$M_1=0.787 \pm 0.012\,M_{\odot}$, $R_1=0.788 \pm 0.015\,R_{\odot}$,
$M_2=0.770 \pm 0.009\,M_{\odot}$, and $R_2=0.817 \pm
0.010\,R_{\odot}$.  The stars in J0516 have radii that are
significantly larger than model predictions for their masses, similar
to what is seen in a handful of other well-studied low-mass
double-lined eclipsing binaries.  We compiled all recent mass and
radius determinations from low-mass binaries and determine an empirical
mass-radius relation of the form
$
R(R_{\odot})=0.0324 + 0.9343M(M_{\odot}) +0.0374M^2(M_{\odot})
$.

\end{abstract}


\section{Introduction}

\begin{table*}[t]
 \caption[Mass and Radius of other double-lined binaries]
 {Mass and Radius of sample double-lined eclipsing binaries plotted in
   Figure \ref{mr}. \label{tab_mr}}
 \begin{tabular}{lccl}    \hline \hline
  Name &  Mass (\Msun) & Radius (\Rsun) & Reference \\  \hline
V818 Tau B & $0.7605 \pm 0.062$ & $0.768 \pm 0.010$ & \citet{tor02} \\
RX J0239.1-1028 A & $0.736 \pm 0.009$ & $0.735\pm 0.018$ &\citet{rib05}\\
RX J0239.1-1028 B &  $0.695 \pm  0.006$ &   $0.710 \pm
0.016$&\citet{rib05}\\
GU Boo A & $0.610 \pm  0.007$ &  $0.623 \pm 0.016$ &\citet{lop04}\\
GU Boo B & $0.599 \pm  0.006$ &  $0.620 \pm   0.020$& \citet{lop04} \\
YY Gem AB & $0.5992 \pm 0.0047$ &  $0.6191 \pm  0.0057$ &\citet{tor02} \\
NSVS0103 A & $0.540 \pm 0.002$ & $0.527 \pm 0.002$ & Lopez-Morales et al.
(in prep)\\
NSVS0103 B & $0.498 \pm 0.002$ & $0.505 \pm 0.002$ & Lopez-Morales et al.
(in prep)\\
TrES-Her0-07621 A & 0.493$\pm$0.003 & 0.453$\pm$0.060 & Creevey et al.
(2005)\\
TrES-Her0-07621 B & 0.489$\pm$0.003 & 0.452$\pm$0.050 & Creevey et al.
(2005)\\
BW3 V38 A & 0.44$\pm$0.07 & 0.51$\pm$0.04 & \citet{mac04}\\
BW3 V38 B & 0.41$\pm$0.09 & 0.44$\pm$0.06 & \citet{mac04}\\
CU Cnc A & $0.4333 \pm 0.0017$ &  $0.4317 \pm  0.0052$&\citet{rib03}\\
CU Cnc B & $0.3890 \pm  0.0014$ &  $0.3908 \pm  0.0094$&\citet{rib03}\\
CM Dra A & 0.2307$\pm$0.0010 & 0.2516$\pm$0.0020 & Lacy (1977) and
Metcalfe et al. (1996) \\
CM Dra B & 0.2136$\pm$0.0010 & 0.2347$\pm$0.0019 & Lacy (1977) and
Metcalfe et al. (1996) \\
\hline
\end{tabular}
\end{table*}

Binary stars offer the best opportunity for accurate measurements of the radii
 and masses of stars, measurements essential to verify stellar evolution
 theory and to determine the properties of other diverse objects such as white
 dwarfs, neutron stars, black holes, and extra-solar planets. These
 measurements rely on dynamical constraints from observed radial and
 rotational velocities, and geometric constraints
 from photometric time series
 observations. Using computer models of binary stars we can
 derive physical parameters (e.g. masses, radii, etc.).
 Understanding the structure and evolution of stars is a basic goal of stellar
 astronomy, and is  required in most other branches of astronomy. Critical 
tests of evolution theory for stars other than the Sun can be made on a small 
set of eclipsing binary stars [see, Pols et al.\ 1997; 
Schr\"oder et al.\ 1997]. 
In general, when accurate tests are available, the results of stellar
 evolution models compare favorably to data for main sequence stars with 
masses greater than one solar mass \citep{pol97}. In contrast, evidence 
has been growing that the models for stars on the lower main sequence have 
problems when confronted with precise data from eclipsing binaries. 
For example, 
\citet{tor02} 
showed that all available evolutionary models underestimate the radii of 
the components of the M-star 
YY Gem by about 20$\%$ and overestimate the effective 
temperatures by 150 K or more. Similar discrepencies are found in V818 Tau 
\citep{tor02}, in CU Cnc 
\citep{rib03}, and GU Boo \citep{lop04}.  See Table \ref{tab_mr}
for a compilation of recent mass and radius determinations derived from
low-mass eclipsing binaries.

The disagreement between the models and the data for 
these  binaries is troubling since models for low mass stars are
used to estimate the ages for open clusters and individual T Tauri
stars by placing them in an HR diagram.  Since the number of well-studied
low-mass binaries is still relatively small, 
observations of additional low mass binaries would be extremely useful.  
The star known as 2MASS J05162881+2607387 (hereafter J0516) was
discovered to be an eclipsing binary by Schuh et al.\ (2003),
who noticed its variability during the course of an extensive
campaign to monitor a pulsating white dwarf.  The eclipse period was
determined to be 1.29395(25) days.  On the basis of a moderate
resolution spectrum and photometric colors, Schuh et al.\
(2003)  determined
an effective temperature of 4200 K for the primary, corresponding to a
spectral type of K7V.  Assuming that the orbital period is the same as
the eclipse period, Schuh et al.\ (2003) 
modelled the light curve and derived
a mass ratio of $M_2/M_1 \approx 0.11$, a radius ratio of $R_2/R_1
\approx 1$, and a temperature ratio of $T_2/T_1 \approx 0.6$.  Based on
their modelling results, Schuh et al.\ (2003) 
suggested that J0516 consists of
a late K-type star paired with a brown dwarf.  Schuh et al.\
(2003) also
briefly considered the possibility that the orbital period of J0156
was twice the eclipse period (i.e.\ 
$2\times 1.29395 = 2.58790$ days). In that case,
the primary and secondary eclipses are identical in their phased light
curve.  Our spectroscopic results
clearly indicate that the
orbital period of J0516 is in fact 2.58791 days.

We discuss below our  observations of J0516, models of the light
and velocity curves, and implications of our results.

\begin{figure*}[t]
\center
\includegraphics[angle=0,scale=.55]{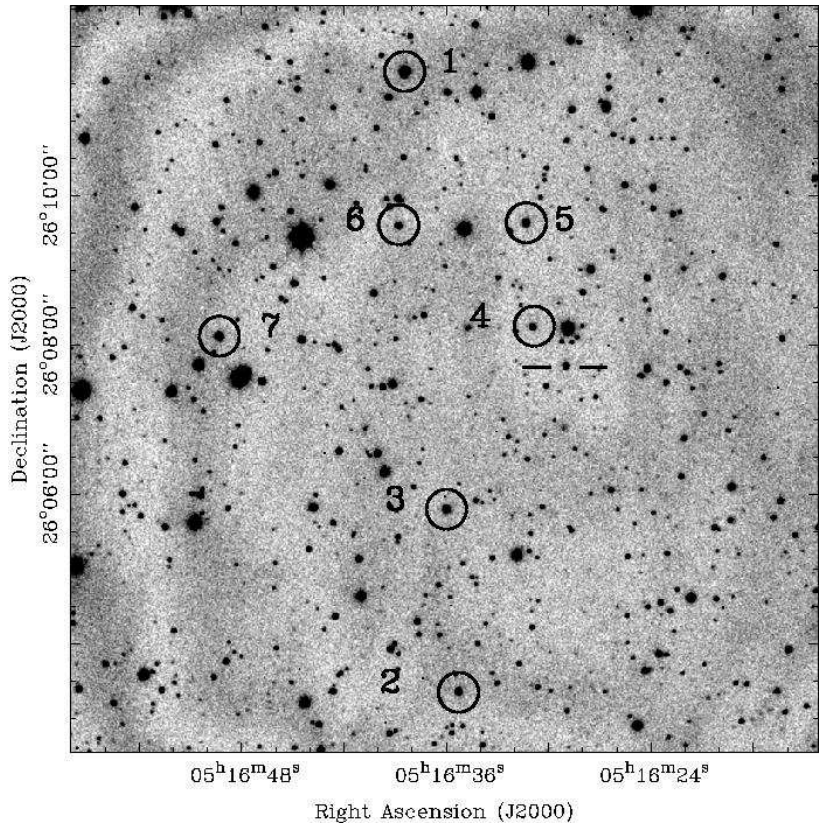}
\caption{
A $10\times 10$ arcminute $I$-band image of the field of J0516
showing
the seven comparison stars (Table 2).
The position of J0516 is indicated by the hash marks.
\label{comp}}
\end{figure*}

\section{Observations and Reductions}

\subsection{Differential Photometry}

\begin{figure*}[t]
\center
\includegraphics[angle=270,scale=.6]{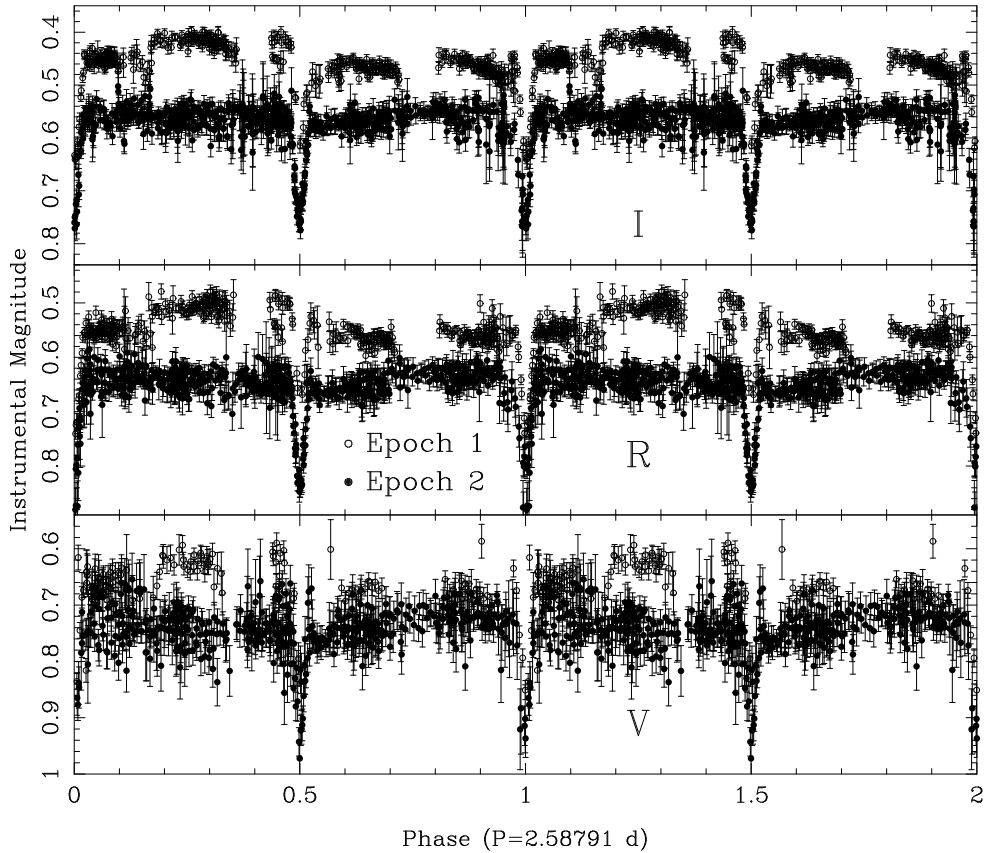}
\caption{The light curves from the two epochs.  The star was more active in 
epoch 1 when compared to the ``quiesent'' period seen in  epoch 2, namely
the star was approximately 0.1 magnitude brighter in all three bandpasses.
\label{lc}}
\end{figure*}

\begin{table}[h]
\center
 \caption[Comparison stars]{USNO-B1 Catalouge Coordinates
   of the seven 
   comparison stars.  The star number indicates the star in Figure 
   \protect{\ref{comp}}
   \label{tab_comp}}
 \begin{tabular}{cccc}    \hline \hline
  Star &  USNO Number & $\alpha$  & $\delta$
 \\  
  &   (U1125) & (J2000) & 
(J2000) \\  \hline
1 & 02245259 & 05 16 38.302 & 26 11 34.98 \\
2 & 02244488 & 05 16 35.311 & 26 03 18.40 \\
3 & 02244635 & 05 16 35.933 & 26 05 44.20 \\
4 & 02243339 & 05 16 30.797 & 26 08 09.85 \\
5 & 02243426 & 05 16 31.133 & 26 09 33.66 \\
6 & 02245357 & 05 16 38.688 & 26 09 32.18 \\
7 & 02248141 & 05 16 49.409 & 26 08 03.88 \\ \hline
 \end{tabular}
\end{table}

J0516 was observed extensively over 44 nights from January 2004 to
January 2006 at the Mount Laguna Observatory (MLO, located in Southern
California).  We used the 1m telescope equipped with a Loral
$2048\times 2048$ CCD (binned $2\times 2$, yielding a pixel scale of
0.8 arcseconds per pixel) and standard $V$, $R$ and $I$ filters.
Typically, the integration times were 
300-360 seconds, 240-300 seconds, and 180 seconds 
for $V$, $R$, and $I$ bandpasses, respectively,
with each observing session lasting several hours per night.  These
observations were divided into two main epochs
based on the source behavior, January - March 2004
(epoch 1) and October 2004 - January 2005, plus January 2006 (epoch 2).  
In  epoch 1, there are 398 $I$ band images, 369 $R$ band
images, and 137 $V$ band images.  There were fewer $V$ band images because of
the 
difficulty of obtaining quality images through the occasional 
clouds and periods of poor seeing. 
In the epoch 2, there are 678 $I$ band
images, 588 $R$ band images, and 481 $V$ band images.

Standard IRAF\footnote{IRAF is distibuted by the National Optical Astronomy
  Observatory, which are operated by the Association of Universities for
  Research in Astronomy, Inc., under the cooperative agreement with the
  National Science Foundation} tasks were used to remove the electronic 
bias and to perform the flat-fielding corrections.  
The IRAF task `imalign' was used 
to remove the differences in the pixel locations of the stellar images 
and to place all the CCD images on the same relative coordinate system.

Differential light curves were derived using a two step process.
Stable comparison stars were found using an automated script that utilizes
Stetson's
programs DAOPHOT IIE, ALLSTAR, and DAOMASTER
(Stetson 1987; Stetson, Davis, \& Crabtree 1991; Stetson 1992a \& 1992b).
To do this, the point spread function (PSF) was determined from each image
from fits to several isolated stars, and the instrumental magnitudes were
computed using an aperture with a radius of 6 pixels.
We   identified seven stable stars that are relatively bright
and isolated.
The coordinates of these seven stars are given
in Table \ref{tab_comp} and a finding chart is shown in Figure \ref{comp}.
Then, the IRAF task `phot' was used to perform aperture
photometry on J0516 and the seven comparison stars using a sequence of
14 concentric apertures of radius 3-16 pixels.  The IRAF
implimentation of Stetson's curve-of-growth technique 
(Stetson 1990)
was used to
derive optimal instrumental magnitudes corresponding to the largest
aperture.  Finally, the light curve of J0516 was found differentially
by using the average instrumental magnitudes of the seven comparison
stars.  
The phased differential
light curves from both epochs are shown in Figure \ref{lc}.
The mean magnitudes of J0516 determined by Schuh et al.\ (2003)
are $V=18.1\pm 0.1$, $R=16.8\pm 0.3$, and $I=15.84\pm 0.3$.
We note that since the source is very 
red,
the signal-to-noise is highest in $I$ and lowest in $V$.

\subsection{Echelle Spectroscopy}

We have obtained 18  spectra ($R=15,000$) using the High
Resolution Spectrograph (HRS,
Tull  1998) and the Hobby-Eberly Telescope (HET,
Ramsey et al.\ 1998).
The instrument configuration consisted of a resolving power of
15,000, the central echelle rotation angle, the 316 groove mm$^{-1}$
cross disperser set to give a central wavelength
of 6948~\AA, the 2 arcsecond science fiber, and two sky fibers.
The source was observed by the HET staff in dark skies and good seeing
conditions between 2005 October 10 and December 27.  
The exposure times were 1800 seconds each, split up into two
900 second parts to aid in the removal of cosmic rays.
We also have the
spectrum of the bright K7V star HD 28343
for comparison purposes, as well as ten 
spectra of six radial velocity standard stars
selected by the HET staff from the list of 
Nidever et al.\ (2002).  

\begin{figure}[t]
\center
\includegraphics[angle=0,scale=.43]{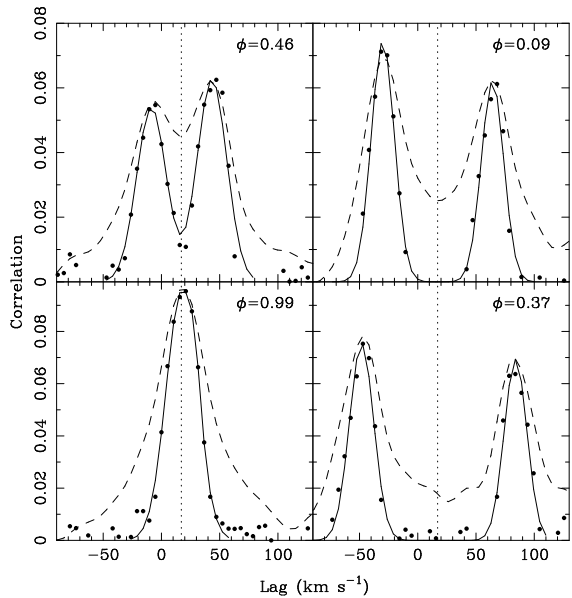}
\caption{Representative broadening functions (BFs)
for four observations
are shown with
filled circles.  Double Gaussian fits to the broadening functions are
shown with the solid lines.  The cross correlation functions (CCFs)
are shown with the dashed lines.  The orbital phases are indicated in the upper
right of each panel, and the vertical dotted lines denote the systemic
velocity of the binary (17.1 km s$^{-1}$).
Note how the BFs have thinner and more separated peaks
than the CCFs.
\label{lag}}
\end{figure}

\begin{figure}[t]
\center
\includegraphics[angle=270,scale=.30]{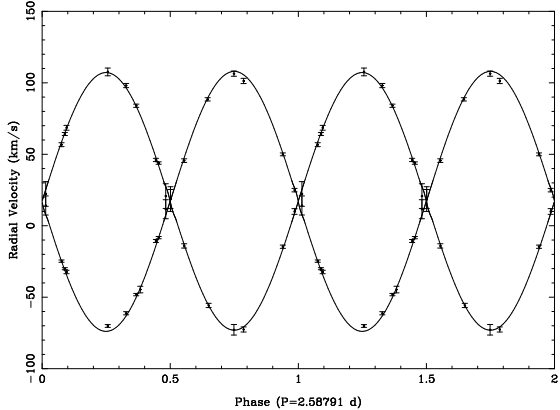}
\caption{The radial velocity curve of the two stars.  \label{rv}}
\end{figure}

The electronic bias was removed from each image.  Then the pairs
of 900 second exposures were combined using the `crreject'
option for cosmic ray removal.  This worked very well, and the resulting
spectra were largely free from cosmic rays.
The echelle spectra were extracted and wavelength calibrated 
using the IRAF `echelle' package.
The signal-to-noise ratios of the extracted
spectra generally lie in the range of
about 10 to 20 per pixel.  
The HRS detector is a
mosiac of two CCDs, and owing to the severe fringing on the ``red''
CCD, we have utilized only the portion of the spectra imaged on the
the ``blue'' CCD, where the wavelength coverage is about 5100-6900
Angstroms.  

The stability of the instrument was checked by cross correlating the spectra
of the standard stars against each other using the `fxcor' task
in IRAF.  The standard deviation of the  differences
between the velocities measured
using fxcor and the velocities given in Nidever et al.\ (2002)
was about 1.7 km s$^{-1}$.

Radial velocities for the J0516 spectra
were derived using the ``Broadening Function''
technique developed by Rucinski (1999).  
This technique for extracting radial velocities of double-lined
binaries is often
more robust than a simple cross correlation
technique (Tonry \& Davis 1979), especially when the velocity
separation between the components is on the order of the spectral resolution.
The spectrum of HD 28343 was used as the high signal-to-noise, sharp-lined
reference spectrum.
In preparation for the analysis, 
each echelle order was normalized to its
continuum level using a three piece cubic spline.  The normalized
spectra were Doppler corrected to the heliocentric rest frame and the 
echelle orders were merged using a linear dispersion
in the wavelength range $5898.68 \le \lambda\le 6689.00$~\AA\
with a pixel size of 0.1168~\AA.  
In fifteen out of the eighteen spectra, the
broadening functions (BFs)
have two clear peaks, indicating that the
spectrum is double-lined.  The remaining three spectra happened to be taken
at conjunction phases, and hence resembled single-lined spectra.
Radial velocities were measured using
Gaussian fits to the peaks in the BFs.  Figure \ref{lag}
shows representative BFs and the double Gaussian fits, and the corresponding
cross correlation functions
(CCFs) for comparison, where the spectrum of HD 28343 was used
as the template spectrum.  
The peaks in the BFs are much better separated than
the peaks in the CCFs, resulting in more reliable radial velocities.
Figure \ref{rv} shows the radial velocities for both components folded
on the period determined below.

The full widths at half maximum
of the Gaussian fits to the J0516 broadening function peaks
average about 16 km s$^{-1}$, compared to an average of 
11 km s$^{-1}$ for the
resulting peaks for the various sharp-lined radial velocity standard
stars.  This indicates that the rotational velocities of the stars in
J0516 are marginally resolved
(e.g.\ the resolving power corresponds to a velocity of
$\approx 20$ km s$^{-1}$), and are on the order of 16 km s$^{-1}$.

\begin{figure*}[t]
\center
\includegraphics[angle=0,scale=.5]{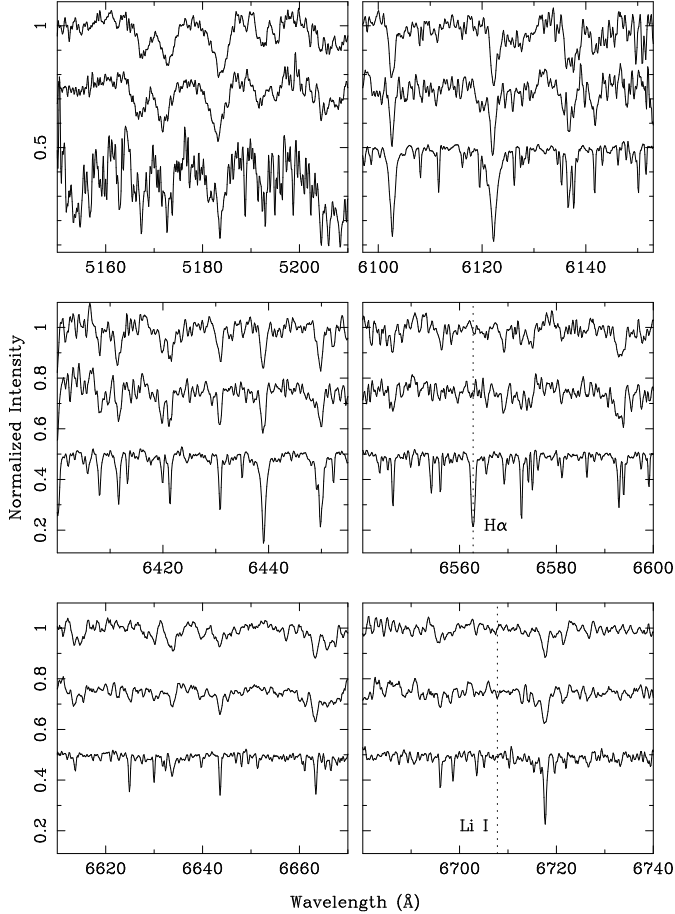}
\caption{The six panels show the smoothed, continuum normalized spectra
of the J0516 components and of the K7V star HD 28343.  In each case,
the spectra from top to bottom are the J0156 primary, the J0516
secondary, offset by 0.25 units, and HD 28343, {\em scaled}
by 0.5 to account for the dilution of the lines in the J0516 spectra.
The J0516 spectra showing the Mg I b feature near 5180~\AA\
have not been sky-subtracted owing to the poor signal-to-noise
there.    
\label{spect}}
\end{figure*}

We constructed ``restframe'' spectra for both stars by Doppler correcting
each of the 18 individual spectra to zero velocity and averaging them
together using `minmax' rejection.  
This was done for the primary, which we define as the component 
that is eclipsed at HJD 2452251.5173 
[the $T_0$ given in Schuh et al. (2003)]
and then again for the secondary.
The lines of the secondary appear to
be removed from the restframe spectrum of the primary (and vice versa) 
reasonably well
using the minmax rejection.  Note, however, that the lines in a
given restframe spectrum are diluted by the continuum from the other
component.
Figure \ref{spect} shows six different
echelle orders of each restframe spectrum.  Each order has been normalized
to its continuum, smoothed using a running average of five pixels,
and the spectrum of the secondary has been offset by
0.25 units for clarity.  
For the sake of the presentation, the J0516 spectra showing the Mg I b 
feature
near 5180~\AA\ have not been sky subtracted owing to the poor signal-to-noise
in the object
there.
Also shown is the spectrum of 
the K7V star HD 28343
(obtained with the same instrumental configuration), 
which has been normalized to its continuum and
scaled by 0.5 in order to account for the dilution of the lines
in the restframe spectra.
Although the signal-to-noise ratios in the combined ``restframe''
spectra are not terribly large, the two stars 
appear to have nearly identical line
features, many of which  resemble the lines seen in the K7V
comparison. Thus the spectral type of K7V determined by
Schuh et al.\ (2003) appears to be correct.
The relative areas under the BFs give the ``luminosity'' ratio, 
$L_2/L_1$, which is unity to within the errors.

There is, however,  very noteable difference between the spectra of 
the J0516 components and the template spectrum: the stars in
J0516 have no significant
feature at H$\alpha$, either in emission  or in
absorption.  
In this regard, J0516 resembles some RS Canum
Venaticorum and BY Draconis binaries that have ``filled in" H$\alpha$ line
profiles (e.g. Fernandez-Figueroa, et al.\ 1994).  
Our
photometric light curves, discussed below, also indicate that one or
both stars in J0516 have relatively high levels of stellar activity.
The equivalent width of the H$\alpha$ absorption line in the K7V template
star is about 0.55~\AA, so presumably each star in J0516 would have to have
an H$\alpha$ emission line with an equivalent width of about 0.55~\AA\
in order to produce a roughly featureless spectrum at H$\alpha$.

We examined the spectra near 6708~\AA\ for an indication of
the Li I doublet, which is sometimes used as an age indicator
(e.g.\ Boesgaard \& Tripicco 1986;
Boesgaard \& Budge 1988).   
There are no significant
lines near this wavelength in either the primary restframe
spectrum or the secondary restframe spectrum, down
to an equivalent width of about
0.02~\AA, which is roughly the noise level (see Figure \ref{spect}).

\begin{figure*}[t]
\center
\includegraphics[angle=270,scale=.68]{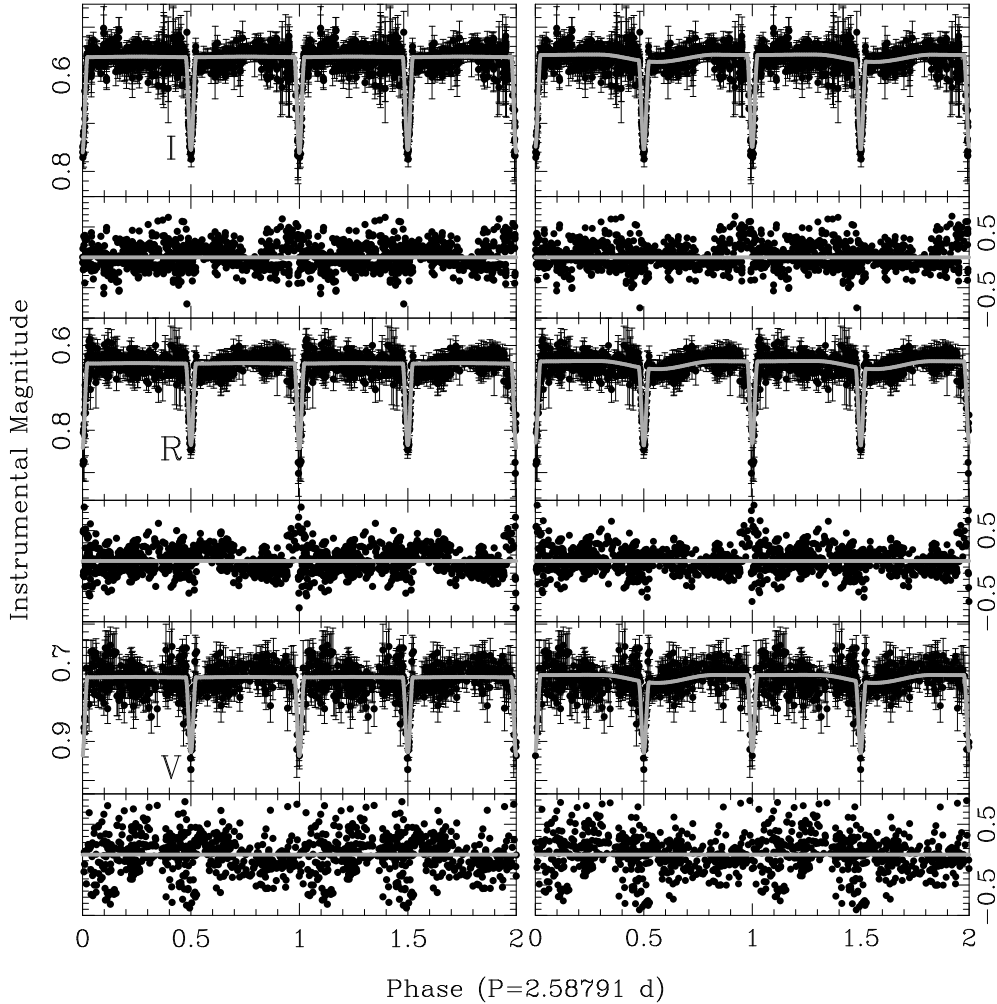}
\caption{The epoch 2 data with the best-fitting model from ELC
and the corresponding residuals.  Left:  the epoch 2 spotless model
and residuals.  Right:  the epoch 2 spot model with a single spot
on each star. 
The bandpasses are, from top to bottom, $I$, $R$, and $V$.
\label{e2lc}} 
\end{figure*}

There are more sophisticated deblending techniques
to analyze the spectra of double-lined binaries, but these typically 
require a grid of template stars.
Thus, we are limited since the only appropriate template spectrum we have
is that of the K7V star HD 28343 (the radial velocity standards observed
for us are all various subtypes of spectral class G).
Nevertheless, we did construct an alternate restframe spectrum
of the primary by scaling the normalized template spectrum by 0.5, 
Doppler shifting
it   to match the velocity of the secondary for each observation, subtracting
it from each observation, Doppler shifting the difference spectra to
remove the velocity of the primary, and finally,
averaging the ``difference spectra'' to produce the final result.  Apart
from having absorption lines a factor of two stronger, this
alternate restframe spectrum looked very similar to the spectrum shown
in Figure \ref{spect}.  The noise levels
in the continua are similar, and there is no feature at H$\alpha$.

\section{Analysis}

\subsection{System Variablity}

The light curves from epoch 1 are distinctly
different than the light curves
from epoch 2.  
As shown in Figure \ref{lc},
the source was roughly 0.1 mag brighter in all
three bandpasses during epoch 1 when compared to epoch 2.  
Although we do not have standard star observations, an inspection
of the instrumental $V-I$ colors indicates that the system was
$\approx 0.02$ mag redder in $V-I$ during epoch 1.
Also,
phased light curves from epoch 1 show considerable out-of-eclipse
variability, presumably due to bright spots on one or both stars.  In
contrast, the phased light curves from epoch 2 show little
out-of-eclipse variability.  Asymmetries in the light curves
seem to be a common feature of low-mass binaries
(e.g.\ the GU Boo light curves shown by Lopez-Morales
\& Ribas 2005).  However, the activity in J0516 seems to be at
a much higher level compared to other well-studied low mass binaries,
especially the $\approx 10\%$ increase on the system brightness 
observed in epoch 1.  If this increase in brightness is due to
bright spots, then one would need to have a substantial portion
of the surface of one or both components covered with bright areas
in order to have them visible at all phases.

\subsection{Light Curve Modeling}

We modeled the light curves using our ELC code
\citep{oro00} with updated model atmospheres for low mass main
sequence stars and brown dwarfs (Hauschildt, private
communication). 
Our ``base''
model has nine free parameters:
the inclination $i$, the mass of the primary $M_1$, the $K$-velocity of the
primary $K_1$, the radius of the primary $R_1$, the temperature of the
primary $T_1$, the ratio of the radii $R_1/R_2$, the ratio of
the temperatures $T_2/T_1$, the orbital period $P$, and the time
of primary eclipse $T_0$.  
Note that for a given orbital period $P$ and inclination $i$, specifying
the primary mass $M_1$ and $K$-velocity $K_1$ uniquely determines the ratio
of masses $q=M_2/M_1$ and the orbital separation $a$.  We have found 
that in cases
such this one, it is more efficient to explore the $M_1,K_1$ parameter
space rather than the $q,a$ parameter space since the observed radial 
velocity curves
give $K_1$ and a rough value of $M_1$ directly.
We adopt a primary temperature of $4200\pm 200$ K \citep{sch03}.
The gravity darkening exponents were set according
to the mean stellar temperatures according to the results of
Claret (2000).  ``Simple'' reflection was used (see Wilson 1990), assuming
bolometric albedos of 0.5 for each star.  We assume a circular orbit
with synchronous rotation for both stars.  We note, finally, that
limb darkening coefficients are not needed since we have model
atmosphere specific intensities tabulated at 99 emergent angles.

The light curves of close binary stars are sometimes asymmetric about
the conjunction phases, and these asymmetries are often attributed to
spots (either bright or dark) on one or both components. 
The spots in ELC are parameterized in the same way as in the
Wilson-Devinney (1971) code.  They are circular regions specified by
four parameters:  the ``temperature factor'' $T_f$, the ``latitude'' of
the spot
center, the ``longitude'' of the spot center, and the angular
radius of the spot.  Bright spots have $T_f>1$ and dark spots have
$T_f<1$.  

We began by modelling the Epoch 2 light curves since these light
curves are better sampled and show very little out-of-eclipse
variability.
The best-fitting model (with no spots)
was arrived at by iteration and brute force.
Our ``observables'' are light curves in $V$, $R$, and $I$,
radial velocity curves for both stars, and a ``luminosity ratio''
of $L_2/L_1=1.00\pm 0.03$, taken to be in the $V$-band.
We ran ELC's genetic optimizer several times to arrive at the best 
intermediate solution
using liberal ranges for the free
parameters.  Using this solution, we scaled the error bars on
the photometry and radial velocities to give $\chi^2=N-1$  for each
data set separately.  The median error bars after this scaling were
$0.030$ mag for $V$, $0.018$ mag for $R$, $0.017$ mag for $I$,
$1.16$ km s$^{-1}$ for the primary radial velocity curve,
and $1.15$ km s$^{-1}$ for the secondary radial velocity curve. 
The data sets with the scaled error bars were optimized again using
the genetic optimizer and a simple ``grid search'' technique.
The best-fitting spotless model
is shown in the left panels of Figure 
\ref{e2lc}.

\begin{table*}[t]
\centering
 \caption[The results of ELC modeling.]{The results of ELC
modeling.
\label{tbl}}
 \begin{tabular}{lccc}    \hline \hline
Parameter         &  Epoch 2 without  & Epoch 2 with  
       & Epoch 1
with \\
               &  spot & spot on each star 
       & 
with spot\\
\hline
Period (days)     &  $2.58791 \pm 0.00001$     & $2.58792\pm 0.00001$ & 
  $2.58791 \pm 0.00002$   \\
$T_0$ (HJD+2,450,000)      & $2251.512 \pm 0.005$ & $2251.508\pm 0.005$ & 
  $2251.517 \pm 0.005$  \\
Inclination (deg) &  $84.1 \pm 0.1$    & $84.3\pm 0.1$ & 
  $84.08 \pm 0.04$        \\
$T_2/T_1$         & $0.989 \pm 0.005$  & $0.988\pm 0.005$ &
        $0.981 \pm 0.004$  \\
$a\, (R_{\odot})$         & $9.19 \pm 0.05$   & $9.19\pm 0.05$ & 
  $9.24 \pm 0.10$          \\
$\log (g_1)$      & $4.51 \pm 0.02$   & $4.54\pm 0.02$ & 
  $4.58 \pm 0.03$         \\
$\log (g_2)$      & $4.49 \pm 0.01$   & $4.50\pm 0.01$ & 
  $4.51 \pm 0.02$         \\
$V_1\sin{i}$ (km s$^{-1}$)     & $15.93 \pm 0.28$  & $15.34\pm 0.30$ & 
  $14.88 \pm 0.50$         \\
$V_2\sin{i}$ (km s$^{-1}$)     & $15.95 \pm 0.19$  & $15.89\pm 0.20$ & 
  $15.88 \pm 0.25$        \\
$K_1$ (\kms)      & $88.42 \pm 0.48$  & $88.45\pm 0.48$ & 
  $88.30   \pm 0.40$        \\
$K_2$ (\kms)      & $90.46 \pm 0.64$  & $90.43\pm 0.60$ & 
  $90.30   \pm 0.60$        \\
$\gamma$ (\kms)   & $17.1 \pm 1.7$    & $17.0\pm 1.7$ & 
  $17.0  \pm 0.1$       \\
$q = M_2/M_1$     & $0.978 \pm 0.019$ & $0.978\pm 0.019$ & 
  $0.978  \pm 0.019$      \\
$M_1$ (\Msun)     & $0.787 \pm 0.012$ & $0.787\pm 0.012$ & 
  $0.786  \pm 0.010$      \\
$M_2$ (\Msun)     & $0.770 \pm 0.009$ & $0.770\pm 0.009$ & 
  $0.769  \pm 0.008$      \\
$R_1$ (\Rsun)     & $0.818 \pm 0.015$ & $0.788\pm 0.015$ & 
  $0.769  \pm 0.015$      \\
$R_2$ (\Rsun)     & $0.820 \pm 0.010$ & $0.817\pm 0.010$ & 
  $0.806  \pm 0.010$      \\
\hline
\end{tabular}
\end{table*}

\begin{figure}[t]
\center
\includegraphics[angle=0,scale=.43]{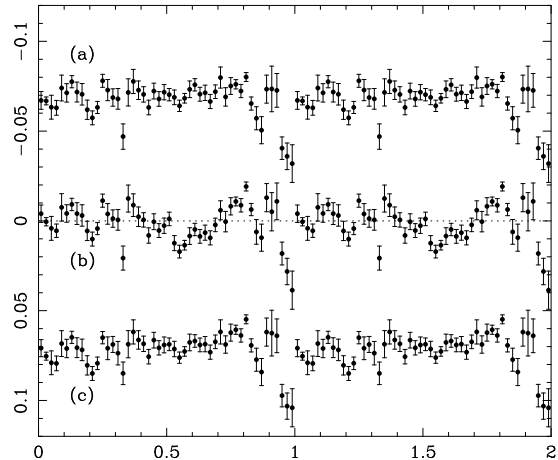}
\caption{$I$-band residuals from epoch 2 model fits are shown
phased on the orbital period and binned into 50 bins.
The residuals marked (a) are from the best-fitting model which had
a single spot on each star.  The residuals marked (b) are for
the model with no spot, and the residuals marked (c) are for
the model with a single spot on the primary (see Tables \ref{tbl} and
\ref{rads} for parameters).  Note the persistent feature near the
primary eclipse (phase 0).}
 \label{plotres}
\end{figure}

We used the procedure outlined in \citet{oro02} to find approximate
1$\sigma$, 2$\sigma$, and 3$\sigma$ confidence intervals.  To estimate
uncertainties on fitted and derived parameters we projected the
9-dimensional $\chi^2$ function into each parameter of interest.  The
1$\sigma$, 2$\sigma$, and 3$\sigma$ confidence limits taken to be the
ranges of the parameter where $\chi^2 \leq \chi^2_{\rm min} + 1$,
$+ 4$, and $+ 9$, respectively.  
Since
the genetic ELC code samples parameter space near $\chi^2_{\rm min}$
extensively, computing these limits is simple.  ELC saves from every
computed model the $\chi^2$ of the fit, the value of the free
parameters, and the astrophysical parameters (e.g., the primary star
mass, the radii of the components, etc.).  One can then select out the
lowest $\chi^2$ at each value of the parameter of interest.  
The values and their uncertainties of various fitting and
derived parameters for the best-fitting Epoch 2 spotless model are given
in Table \ref{tbl}.

We then checked for systematic trends by searching for periodicities
in the $I$-band residuals (the $I$-band light curve has the best 
signal-to-noise).   The Lomb-Scargle periodogram 
(Lomb 1976; Scargle 1982)
showed significant
power at the orbital period.  Indeed, a small modulation with
an amplitude of about 1\% is evident in the $I$-band residuals
when they are
phased on the orbital period and binned into 50 phase bins
(see Figure \ref{plotres}).  There is also a small systematic
problem with the model near primary eclipse.  

\begin{figure*}[t]
\center
\includegraphics[angle=270,scale=.5]{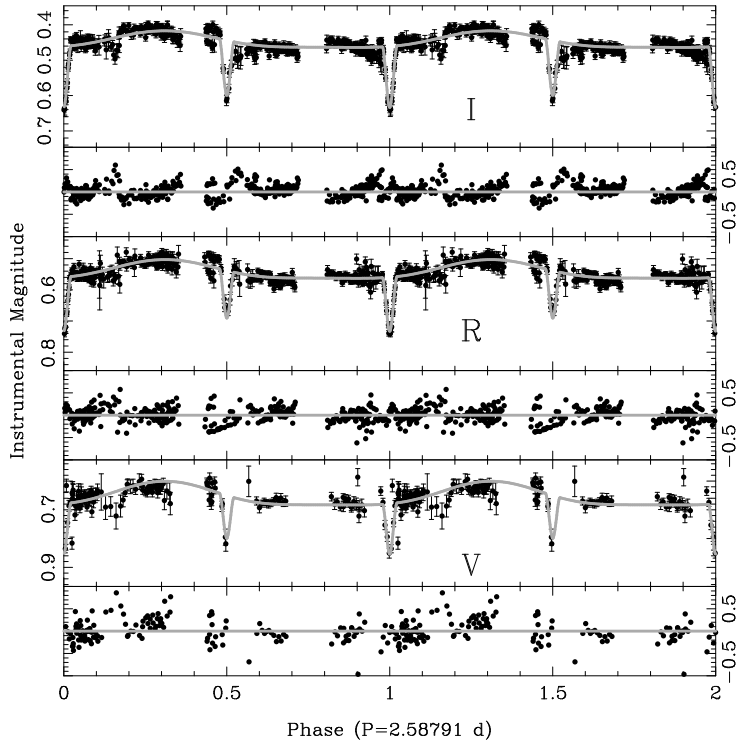}
\caption{The epoch 1 data with the best-fitting model from ELC
and the corresponding residuals.  See Table \ref{tbl} for the parameters.
The bandpasses are, from top to bottom, $I$, $R$, and $V$.
}
 \label{E1lc}
\end{figure*}

We performed additional fits to the Epoch 2
light curves using models with spots.  We had five different spot
configurations:  a spot on the primary only, a spot on the secondary
only, two spots on the primary only, two spots on the secondary only,
and a single spot on both the primary and secondary.  All of the models
with spots had $\chi^2$ values that are significantly lower
than the spotless model (see Table \ref{rads}).
The phased and binned $I$-band residuals for the spotted
models look very similar to each other.    Figure \ref{plotres}
shows the phased and binned residuals for two of the spot models.
The low-level
modulation near phase 0.5 is gone, but the feature near primary eclipse
at phase 0 remains.
Apart from some small differences in the derived radii of the stars,
the model and derived parameters for the spotted models agree
quite well with the spotless model.  As an example,
Table \ref{tbl}
gives the parameters for the model with a single spot on each star,
and the panels on the right side of Figure  \ref{e2lc} show this model.

The temperature of the primary is not constrained by our light curves
alone.  However, the ratio of the temperatures is quite well constrained
by the light curves, and is quite close to unity (see Table \ref{tbl}),
as we expected based on the spectroscopic results (i.e.\ the restframe
spectrum of the primary  looks the same as
the restframe spectrum of the secondary to within the noise level).

In all of the modelling described previously, we used  the observed
 ``luminosity ratio'' in the $V$-band of $1.0\pm 0.03$ as an additional
constraint.  Basically, we compute the quantity $\chi^2_{\rm lum}
= (1-L_2/L_1)^2/0.03^2$ and add it to the total $\chi^2$,
which has the effect of selecting models that have
$L_2\approx L_1$.
This is an entirely reasonable constraint since the spectroscopic
results indicate that the primary star has the same spectral type
as the secondary
star, and that the two stars have nearly equal masses (i.e.\
$K_1\approx K_2$).  In order to test for possible systematic errors,
we did additional model fits to the epoch 2 light curves without
the  luminosity ratio constraint.  For the baseline model with no spots,
we found basically the same temperature ratio  ($T_2/T_1=0.984$) as
before.  However, the ratio of the radii was
$R_1/R_2=1.064$, which yields a $V$-band luminosity ratio of
$L_1/L_2\approx 1.5$.  It seems unlikely that two stars with nearly
equal masses and temperatures should have such dissimilar radii,
so we reject this model and only  consider the models that
have the luminosity ratio constraint imposed.

The most interesting astrophysical parameters are the masses and
radii of the stars, and Table \ref{rads} gives these values for
all of the Epoch 2 models.  Formally, the model with a single spot on each
star gives the best fit.  
While the derived masses depend mainly on the radial velocity curves,
the  light curves have some weak dependence
on the scale of the binary
since ELC uses model atmosphere intensities that are tabulated in
temperatures and gravities.   As a result, the derived masses for each of
the models are not exactly the same.  There is  a maximum spread of
$0.007\,M_{\odot}$ for the mass of the primary and
$0.006\,M_{\odot}$ for the secondary, respectively, which is
is  less than the formal $1\sigma$ errors.  Based on the
best-fitting model, we
adopt values of $M_1=0.787\pm 0.012\,M_{\odot}$
and $M_2=0.770\pm 0.009\,M_{\odot}$ for the primary and secondary masses,
respectively.
On the other hand, the derived radii show a bit more spread between
the various models.  There is a maximum spread of
$0.020\,R_{\odot}$ for the radius of the primary and
$0.013\,R_{\odot}$ for the radius of the secondary, respectively.
Based on the  best-fitting model, we adopt values
of
$R_1=0.788\pm 0.015\,R_{\odot}$ and
$R_2=0.817\pm 0.010\,R_{\odot}$ for the primary and secondary radii,
respectively.  We caution that these values for the radii probably have
systematic errors on the order of $0.02\,R_{\odot}$, given the spread
in radii for the various models, the persistent feature in the
residuals near phase 0 shown in Figure \ref{plotres},
and the need to impose the spectroscopically determined
luminosity ratio.

{
\begin{table*}[t]
\caption{Masses and radii for Epoch 2 models}\label{rads}
\begin{tabular}{lcccccl}
model & $M_1$ & $M_2$ & $R_1$ & $R_2$ & $\chi^2$ & comment \\
\hline
no spots & $0.787\pm 0.012$ & $0.770\pm 0.009$ & 
   $0.818\pm 0.015$ & $0.820\pm 0.010$ & 1774 & \nodata \\
1 spot each & $0.787\pm 0.012$ & $0.770\pm 0.009$ & 
   $0.788\pm 0.015$ & $0.817\pm 0.010$ & 1551 & dark and bright spot \\
2 spots, primary & $0.792\pm 0.012$ & $0.774\pm 0.009$ & 
   $0.797\pm 0.015$ & $0.819\pm 0.010$ & 1570 & dark spots \\
2 spots, secondary & $0.787\pm 0.012$ & $0.769\pm 0.009$ & 
   $0.795\pm 0.015$ & $0.823\pm 0.010$ & 1576 & bright spots \\
1 spot, primary & $0.787\pm 0.012$ & $0.770\pm 0.009$ & 
   $0.797\pm 0.015$ & $0.810\pm 0.010$ & 1603 & dark spot \\
1 spot, secondary & $0.785\pm 0.012$ & $0.768\pm 0.009$ & 
   $0.814\pm 0.015$ & $0.818\pm 0.010$ & 1604 & dark spot \\
epoch 1 & $0.786\pm 0.010$ & $0.769\pm 0.015$ & 
   $0.769\pm 0.015$ & $0.806\pm 0.010$ & \nodata & bright spot \\
\hline
\end{tabular}
\end{table*}
\normalsize}

For completeness, we also fit the epoch 1 data using the same basic
model as above and adding a single spot on the primary.  The parameters
are given in Table \ref{tbl} and the phased light curves,
the
best-fitting models, and residuals are shown in Figure \ref{E1lc}.  The
model fits reasonably well, and for the most part the fitted and
derived parameters agree quite well with those found from the
epoch 2 fits, in spite of the fact that the secondary 
eclipse is not well sampled in epoch 1,
and, as noted above, there is considerable out-of-eclipse variability
in epoch
1.
The one noteable exception is the primary radius,
which is $0.769\pm 0.015\,R_{\odot}$ from the epoch 1 model and
$0.788\pm 0.015\,R_{\odot}$ from the epoch 2 model, a difference
of nearly $0.02\,R_{\odot}$.  Although the
epoch 1 residuals have more scatter than their epoch 2 counterparts,
the epoch 1 residuals show no significant power at
the orbital period in a Lomb-Scargle periodogram.

\subsection{Eclipse Timings}

\begin{figure}[t]
\center
\includegraphics[angle=270,scale=.3]{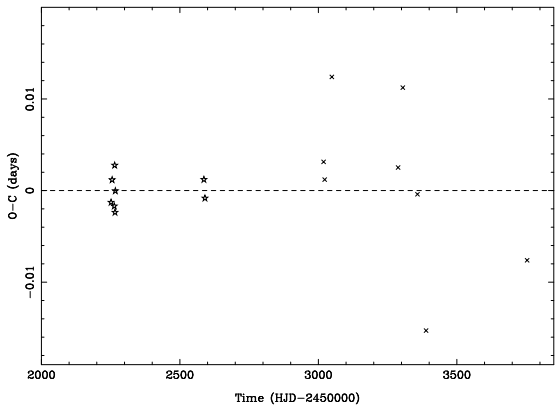}
\caption{O-C diagram of J0516.  The Xs indicate data from this work.  
The stars indicate data from \citet{sch03}. 
\label{oc}}
\end{figure}

\begin{table*}[t]
\centering
 \caption[Eclipse times and O-C]{The HJD of the eclipse times, 
epochs, and O-C values.
\label{timeoc}}
 \begin{tabular}{cccc}    \hline \hline
Eclipse Time (HJD) & Epoch & $O-C$ (days) & Reference \\ \hline 
2452251.5164 & 0.0 &  $-0.0013$ & \citet{sch03}\\
2452255.4007 & 1.5 &  $0.0012$ & $\shortparallel$\\
2452263.1615 & 4.5 & $-0.0017$ & $\shortparallel$\\
2452264.4599 & 5.0 &  $0.0028$ & $\shortparallel$\\
2452265.7487 & 5.5 & $-0.0024$ & $\shortparallel$\\
2452267.0450 & 6.0 & $0.0000$ & $\shortparallel$\\
2452586.6506 & 129.5 & $0.0012$  & $\shortparallel$\\
2452590.5304 & 131.0 & $-0.0008$ & $\shortparallel$\\
2453018.8301 & 296.5 & $0.0031$  & This work\\
2453022.7100 & 298.0 & $0.0012$  & $\shortparallel$\\
2453048.6001  & 308.0 & $0.0124$  & $\shortparallel$\\
2453287.9700 & 400.5 & $0.0025$  & $\shortparallel$\\
2453304.8000  & 407.0 & $0.0112$ &  $\shortparallel$\\
2453357.8401 & 427.5 & $-0.0004$  & $\shortparallel$\\
2453388.8799 & 439.5 & $-0.0153$ & $\shortparallel$\\
2453753.7800 & 580.5 & $-0.0076$  & $\shortparallel$\\ \hline
\end{tabular}
\end{table*}

We determined
eclipse times by fitting a parabola to the $I$-band observations
near times of eclipse (the $I$-band observations have the highest
signal-to-noise).  
The timings and cycle counts are given in Table \ref{timeoc} and have
typical uncertainties of about 10 minutes.
The table also gives the eclipse times determined by 
\citet{sch03} with updated cycle counts.  \citet{sch03}
give no
uncertainties in their eclipse timings, but from an inspection of
their plotted light curves, it seems likely that their timings are
more accurate than ours, so
we assign their timing three times more
weight than our timings. 
These timings are described by the linear ephemeris  
\begin{displaymath}
{\rm min~I} = {\rm HJD}\, 2,452,251.518(1)+2.587890(7)E
\end{displaymath}
(error in the last digit is in parenthesis).  This period differs by
$\approx 2\sigma$ from the spectroscopic/photometric period derived from
the ELC modelling.
An $O-C$ diagram  is shown in Figure \ref{oc}.
The maximum deviation is on the order of 20 minutes, and there is 
no obvious trend.

\begin{figure*}[t]
\center
\includegraphics[angle=270,scale=.6]{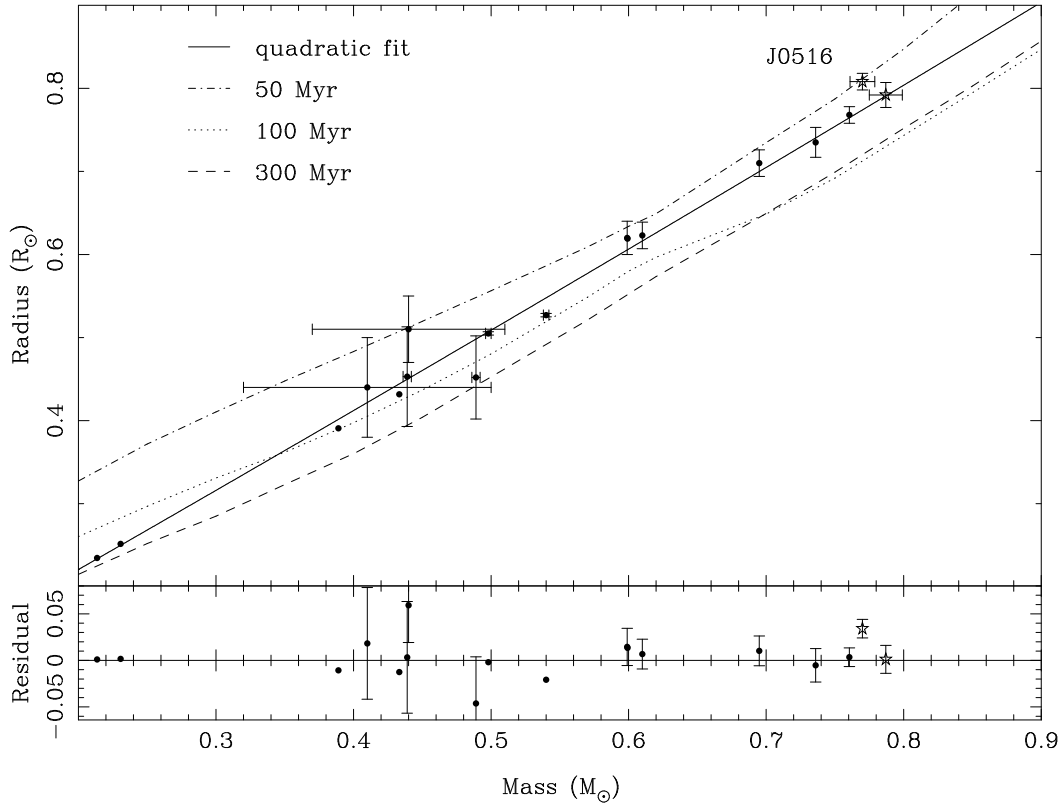}
\caption{Top:
Mass versus radius of some observed ``double-lined'' eclipsing
  binaries mentioned in the text.  J0516 is repesented by the two stars.
Error bars smaller than the symbol size have been omitted.
  Evolutionary models from \citet{bar02}
with ages of 50 Myr, 100 Myr, and 300 Myr are
represented by the dot-dashed line, dotted line, and dashed lines, 
respectively.  
The solid line is weighted a quadratic fit to the
observed data (including the Sun and excluding J0516).  
The secondary of J0516 has a larger radius than what
 the observed relationship defined by the other stars predicts. 
Bottom: The residuals from the quadratic fit.
\label{mr}}
\end{figure*}

\section{Summary and Discussion}

We have derived fairly accurate values of the masses and radii of the
two stars in J0516.
The masses are determined to
1.5\% precision
and the radii are formally determined to 
1.9\% precision, although there may be a systematic error of a few percent
on the radius determinations.
This high level of precision compares
favorably to the precision achieved on the small number
of other low-mass eclipsing binaries mentioned in Section 1.  
For low-mass stars with well determined masses and radii,
the observed radii are consistently larger than what evolutionary models
predict for their masses.  
The stars in J0516 are no exception. 
Figure \ref{mr} shows the stars in J0516 and those of other well-studied
low-mass
binaries in a mass-radius plot.  We also show in Figure
\ref{mr}  theoretical mass-radius relationships with ages of 50 Myr,
100 Myr, and 300 Myr 
taken from the models of Baraffe et al.\ (2002).
No single model passes through all of the observations, and all of the
stars are above the 300 Myr model.
We 
formed an empirical mass-radius relationship by fitting a parabola
using weights on both the mass and radius
through the other stars and the Sun (excluding J0516)
and found the following equation
$$
R(R_{\odot})=0.0324 + 0.9343M(M_{\odot}) +0.0374M^2(M_{\odot}).
$$
Compared to this empirical relationship, the J0516 primary agrees quite well,
whereas the J0516 secondary is slightly too large.

Why is the secondary of J0516 so large for is mass?
Ribas (2005) raises the possibility that the high level of
stellar activity observed in the well-studied double-lined
low-mass eclipsing binaries is what causes these stars
to have radii that are larger than what is predicted.  As we
discussed earlier, J0516 is much more active than the other binaries,
and the secondary's
deviation from the mass-radius relationship is the most extreme.
It is not clear, however,  if higher levels of
activity would automatically lead to larger  radii.

The Li I line is often used as an age indicator.
An upper limit to the equivalent width of the Li I line at 6708~\AA\
of 20 m\AA\ translates roughly to an abundance of
$A({\rm Li})=0.0$,
which translates into a lower limit on the age of the binary
of about 150 Myr (Steinhauer 2003; Deliyannis  private communication).
The evolutionary models of Baraffe et al.\ (2002) indicate that
a $0.8\,M_{\odot}$ star takes about 70 Myr to contract and reach its
``normal'' main sequence radius.  Taken at face value, the lower limit
on the age of 150 Myr from the lack of lithium
indicates that J0516 is well past its pre-main
sequence phase, and that the unusually large radius of the secondary
cannot be explained by youth.

The only practical way to establish the age of J0516 would be to 
associate it with a cluster or moving group.  To do this, one
would need good  proper motions and distance for J0516 and a large
number of stars in the nearby field.  We know of no 
measurement of the proper motion of J0516 [for example, no
proper motion measurement is given in the USNO-B catalog 
(Monet et al.\ 2003)].  
The distance can be estimated from
our model, but the result depends on the extinction and on the
assumed temperatures of the stars.   
Schuh et al.\ (2003) determined a color excess of $E(B-V)=0.9\pm 0.2$ mag
from their model of the spectral energy distribution.
Assuming $A_V=3.1E(B-V)$ and $A_K=0.114A_V$
(Cardelli, Clayton, \& Mathis 1989), the $K$-band extinction is 
$0.318\pm 0.017$ mag.
The absolute $K$ magnitude of the binary was found 
using filter-integrated
surface brightnesses computed from {\sc NextGen} models by
France Allard
(private communication),   assuming the radii given in Table \ref{tbl} and
temperatures of $4200\pm 200$ K. 
The average apparent magnitude in the $K$-band
determined by the 2MASS survey
is $K=13.113\pm 0.039$ (see Schuh et al.\ 2003).
We find a distance of
$d=753 \pm 34$ pc. This distance is substantially
larger than the distances to the other well-studied low-mass
binaries mentioned earlier [e.g.\ $d=140\pm 8$ pc in the case of
GU Boo \citep{lop04}].

J0516 is clearly a very important system in that it allows for further
observational constraints for the lower main sequence.  Owing to the
enhanced variability, it would be worthwhile to obtain additional photometry
to better define the ``quiescent'' phase and to better establish
the eclipse profiles in quiescence.  In addition,
moderate resolution but high signal-to-noise spectroscopy would be
useful in order to further 
investigate the apparent lack of an H$\alpha$ feature
in either star and to check for possible correlations between the
profile of the H$\alpha$ feature (if any)
and the activity level.

\acknowledgments
The Hobby-Eberly Telescope (HET) is a joint project of the University
of Texas at Austin, the Pennsylvania State University, Stanford
University, Ludwig-Maximilians-Universit\"at Munchen, and
Georg-August-Universit\"at G\"ottingen. The HET is named in honor of its
principal benefactors, William P. Hobby and Robert E. Eberly.
This research has also
made use of the NASA/IPAC 
Infrared Science Archive, 
which is operated by the Jet Propulsion 
Laboratory, California Institute of Technology, 
under contract with the National Aeronautics and Space Administration,
and also NASA's Astrophysics Data System.
We  thank Danielle Martino for the data taken at MLO from
January - March 2004 and Con Deliyannis for helpful discussions.

Facilities: \facility{Mount Laguna Observatory,
Hobby Eberly Telescope}

\end{document}